\documentclass{emulateapj}
\newcommand{\msun}{\ensuremath{\rm M_\odot}}
\newcommand{\msunyr}{\ensuremath{\rm M_{\odot}\;{\rm yr}^{-1}}}
\newcommand{\Ha}{\ensuremath{\rm H\alpha}}

\newcommand{\lya}{\ensuremath{\rm Ly\alpha}}

\newcommand{\kms}{km~s\ensuremath{^{-1}\,}}
\newcommand{\ztwo}{\ensuremath{z\sim2}}

\begin{document}

\title{A MODEL FOR STAR FORMATION, GAS FLOWS AND CHEMICAL
  EVOLUTION\\ IN GALAXIES AT HIGH REDSHIFTS}
\author{\sc Dawn K. Erb}
\affil{Harvard-Smithsonian Center for Astrophysics, 60 Garden Street,
  Cambridge, MA 02138} 
\email{derb@cfa.harvard.edu}

\shorttitle{STAR FORMATION, GAS FLOWS AND CHEMICAL EVOLUTION}
\shortauthors{D. K. ERB}

\slugcomment{Accepted for publication in ApJ}

\begin{abstract}
Motivated by the increasing use of the Kennicutt-Schmidt (K-S) star
formation law to interpret observations of high redshift galaxies, the
importance of gas accretion to galaxy formation, and the recent
observations of chemical abundances in galaxies at \ztwo--3, I use
simple analytical models to assess the consistency of these processes
of galaxy evolution with observations and with each other.  I derive
the time dependence of star formation implied by the K-S law, and show
that the sustained high star formation rates observed in galaxies at
\ztwo--3 require the accretion of additional gas.  A model in which
the gas accretion rate is approximately equal to the combined star
formation and outflow rates broadly reproduces the observed trends of
star formation rate with galaxy age. Using an analytical description
of chemical evolution, I also show that this model, further
constrained to have an outflow rate roughly equal to the star
formation rate, reproduces the observed mass-metallicity relation at
\ztwo.

\end{abstract}
 

\section{Introduction}
The motivations of this paper are several.  First, the empirical
Kennicutt-Schmidt (K-S) law \citep{s63,k98schmidt}, which states that
the surface density of star formation is proportional to the surface
density of gas, is widely used to interpret and describe star
formation in galaxies, though its origins are not fully understood and
it is just beginning to be tested at high redshift
\citep{btg+04,csn+07,bcd+07}.  The K-S law is a valuable tool when gas
masses are not directly measurable; it has been used to estimate the
gas masses of both high redshift galaxies \citep{ess+06mass} and local
galaxies in the distant past \citep{cjp+07}.  The consequences of the
K-S law for the evolution of star formation at high redshift are
therefore worth considering in more detail, as is the consistency of
these consequences with observations.

Second, the fueling of star formation by gas accretion is an essential
element of models of galaxy formation, but has largely been neglected
by observers of galaxies at high redshifts, mostly because of the lack
of evidence for inflow in the spectra of these galaxies. In contrast,
the evidence for strong outflows in galaxies at \ztwo--3 is
well-known, most notably in the form of offsets between the redshifts
of nebular emission lines, rest-frame UV absorption lines, and
\lya\ emission (\citealt{pss+01}, Steidel et al.\ 2007, in prep).  In
spite of the lack of observations of inflow, however, its effects
should be considered in the context of other observations, since a
significant inflow rate would affect other, measurable properties.

Finally, in recent years metallicity measurements of increasingly
large samples of galaxies at $z>1$ have become possible
(e.g.\ \citealt{kk00,pss+01,sga+04,sep+04,mlc+06}), including the
detection of a mass-metallicity relation at $z>2$ \citep{esp+06}.
These measurements still suffer considerably from limitations on the
methods that can be used and from calibration uncertainties, but
nevertheless they offer a unique opportunity to place constraints on
star formation histories and gas flows at high redshift, provided that
the effects of inflow, outflow and star formation can be disentangled.
Many recent studies have addressed this issue in some detail,
including \citet{ke99,kh05,d07}; and \citet{fd07}.

The goal of this paper is to formulate simple, analytical models
including star formation according to the K-S law, gas inflows and
outflows, and chemical evolution.  We would like to assess whether or
not these models are consistent with each other and with our current
observational knowledge of high redshift galaxies, and to see if they
might give a general picture of how gas flows, star formation and
metal enrichment may proceed at high redshift.  In \S\ref{sec:ks} I
derive the explicit time dependence of star formation implied by the
K-S law and test its consistency with observations of galaxies at
\ztwo.  In \S\ref{sec:metals} I consider simple models of chemical
evolution which incorporate both inflows and outflows of gas, and again
test their consistency with observations of high redshift galaxies and
with the results of the previous section.  Some implications of the
results are discussed in \S\ref{sec:discuss}.  I adopt a cosmology
with $H_0=70\;{\rm km}\;{\rm s}^{-1}\;{\rm Mpc}^{-1}$, $\Omega_m=0.3$,
and $\Omega_{\Lambda}=0.7$, and use the \citet{c03} initial mass
function (IMF).

\section{The Time Evolution of Star Formation According to the Kennicutt-Schmidt Law}
\label{sec:ks}

The first question is whether or not the locally observed K-S law is
consistent with the observed and inferred masses, star formation rates
(SFRs), and star formation histories of high redshift galaxies.  This
is most easily addressed by deriving the explicit time dependence of
the K-S law, which is written as $\Sigma_{\phi}=C\Sigma_{\rm gas}^n$,
where the star formation rate is $\phi(t)$,
$\Sigma_{\phi}=\phi(t)/r^2$ is the surface density of star formation,
and $\Sigma_{\rm gas}=M_{\rm gas}/r^2$ is the cold gas mass surface
density.  I set $n=1.4$ according to \citet{k98schmidt}, and use
$C=5.5\times10^{-13}$.  For $n=1.4$, the \citet{k98schmidt} value of
$2.5\times10^{-4}$ \msunyr\ kpc$^{-2}$ becomes $1.0\times10^{-12}$
when the gas surface density is given in \msun\ kpc$^{-2}$ rather than
\msun\ pc$^{-2}$, and an additional factor of 1.8 accounts for the
change from Salpeter to Chabrier IMF.  When the K-S law is written as
$\Sigma_{\phi}=C\Sigma_{\rm gas}^n$ and mass is given in \msun, time
in years, and size in kpc, $C$ has units of \msun$^{1-n}$ yr$^{-1}$
kpc$^{2(n-1)}$.  For simplicity I assume that the galaxy size $r$
remains constant with time; this is undoubtedly an oversimplification,
but varying $r$ between reasonable values makes no qualitative
difference to the result.  Neglecting (for the moment) inflows,
outflows, and the gas returned to the ISM by star
formation\footnote{With the inclusion of the gas returned to the ISM
by evolved stars, the gas mass at a given time is described by
$M_{g}=M_{\rm i}-\alpha M_{\star}$, where $M_{\star}$ is the stellar
mass and $\alpha$ is the fraction of the total mass formed into stars
that remains locked in long-lived stars and remnants. The lockup
fraction $\alpha$ is a decreasing function of time, as lower mass
stars evolve off the main sequence, and also depends on the assumed
IMF and on the mass of the remnants.  I assume $\alpha=1$ for
simplicity, but see further discussion at the end of this section and
in \S\ref{sec:metals}.  The issue is also reviewed by \citet{t80} and
\citet{g97}, among others.}, the gas mass is
\begin{equation}
M_{g}(t)=M_{\rm i}-\int_0^t \phi(t')dt', 
\label{eq:gasmass}
\end{equation}
where the initial gas mass is $M_{\rm i}$.  Then the K-S law can be
recast in terms of the SFR, the initial gas mass, and the size as
\begin{equation}
\phi(t)=C \left[M_{\rm i}-\int_0^t \phi(t')dt'\right]^n r^{2(1-n)}.
\end{equation}
In order to solve for $\phi(t)$, we rearrange and isolate the integral:
\begin{equation}
\int_0^t dt^\prime\,\phi(t^\prime)=-\left[\frac{\phi(t)
      }{Cr^{2(1-n)}}\right]^{1/n}+M_i.
\end{equation}
Differentiating both sides and rearranging further leads to the
differential equation
\begin{equation}
\frac{d\phi(t)}{dt}=A\phi(t)^{2-1/n},
\end{equation}
where $A=-n(Cr^{2(1-n)})^{1/n}$.  This has the solution
\begin{equation}
\phi(t)=n^{n/(n-1)} \left[ A\,t(1-n)+n \phi_0 ^{(1-n)/n}\right]^{n/(1-n)},
\label{eq:sfrt}
\end{equation}
where the initial star formation rate $\phi(0)=\phi_0=C M_{\rm
  i}^nr^{2(1-n)}$.   

\begin{figure}[htbp]
\plotone{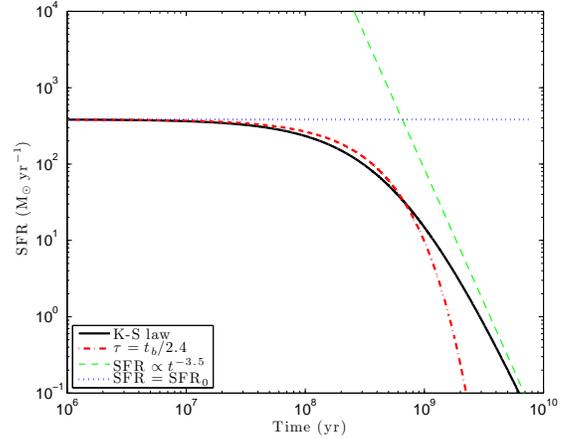}
\caption{The solid black line shows the time evolution of the star
  formation rate given by the K-S law, for an initial gas mass $M_{\rm
    i}=10^{11}$ \msun\ and a fixed galaxy size $r=5$ kpc.  At early
  times the SFR is approximately its initial value (blue dotted line),
  while at late times it falls as $\phi(t)\propto t^{n/(1-n)}$ (green
  dashed line).  The blue and green lines meet at the break time
  $t_b$.  A comparable exponentially declining SFR is shown by
  the red dot-dashed line; in this case $\tau=270$ Myr.}
\label{fig:ks}
\end{figure}

The behavior of this solution is shown in Figure~\ref{fig:ks}, for an
initial gas mass $M_{\rm i}=10^{11}$ \msun\ and a fixed size $r=5$
kpc.  At early times the star formation rate is approximately equal to
the initial SFR, while at late times it falls as a power law with
$\phi(t)\propto t^{n/(1-n)}$.  The early and late solutions meet at a
break time 
\begin{equation}
t_b=\frac{n \phi_0^{(1-n)/n}}{A(1-n)}.
\end{equation}
Regardless of the initial gas mass and size, the star formation rate
at the break time is 9\% of its initial value, and about 80\% of the
final stellar mass has been formed; thus $t_b$ represents a reasonable
parameterization of the lifetime of the galaxy.  For the parameters
shown in Figure~\ref{fig:ks}, $t_b=650$ Myr and $\phi_0=380$ \msunyr;
for reference, values of $t_b$ and $\phi_0$ for other initial gas
masses are given in Table~\ref{tab:sftimes}.  Also note that galaxies
with larger masses have shorter star formation timescales, in
accordance with observations that more massive galaxies form their
stars at earlier times (e.g.\ \citealt{jgc+05,pjhc07}).

\begin{deluxetable}{l c c c}
\tablewidth{0pt}
\tabletypesize{\footnotesize}
\tablecaption{Star Formation Rates and Timescales\label{tab:sftimes}}
\tablehead{
\colhead{$M_{i}$\tablenotemark{a}} &
\colhead{$\phi_0$\tablenotemark{b}} &
\colhead{$t_b$ ($f=0$)\tablenotemark{c}} &
\colhead{$t_b$ ($f=0.9$)\tablenotemark{d}} \\
\colhead{(\msun)} &
\colhead{(\msunyr)} &
\colhead{($10^9$ yr)} &
\colhead{($10^9$ yr)}
}
\startdata
$10^9$ & 0.6 & 4.1 & 41\\
$10^{10}$ & 15 & 1.6 & 16\\
$10^{11}$ & 380 & 0.65 & 6.5\\
$10^{12}$ & 9600 & 0.26 & 2.6\\
\enddata
\tablenotetext{a}{Initial gas mass}
\tablenotetext{b}{Initial star formation rate, assuming $r=5$ kpc}
\tablenotetext{c}{Break time with no inflow or outflow, assuming $r=5$ kpc}
\tablenotetext{d}{Break time with inflow rate 95\% of gas processing rate, assuming $r=5$ kpc}
\end{deluxetable}

\begin{figure*}[htbp]
\plottwo{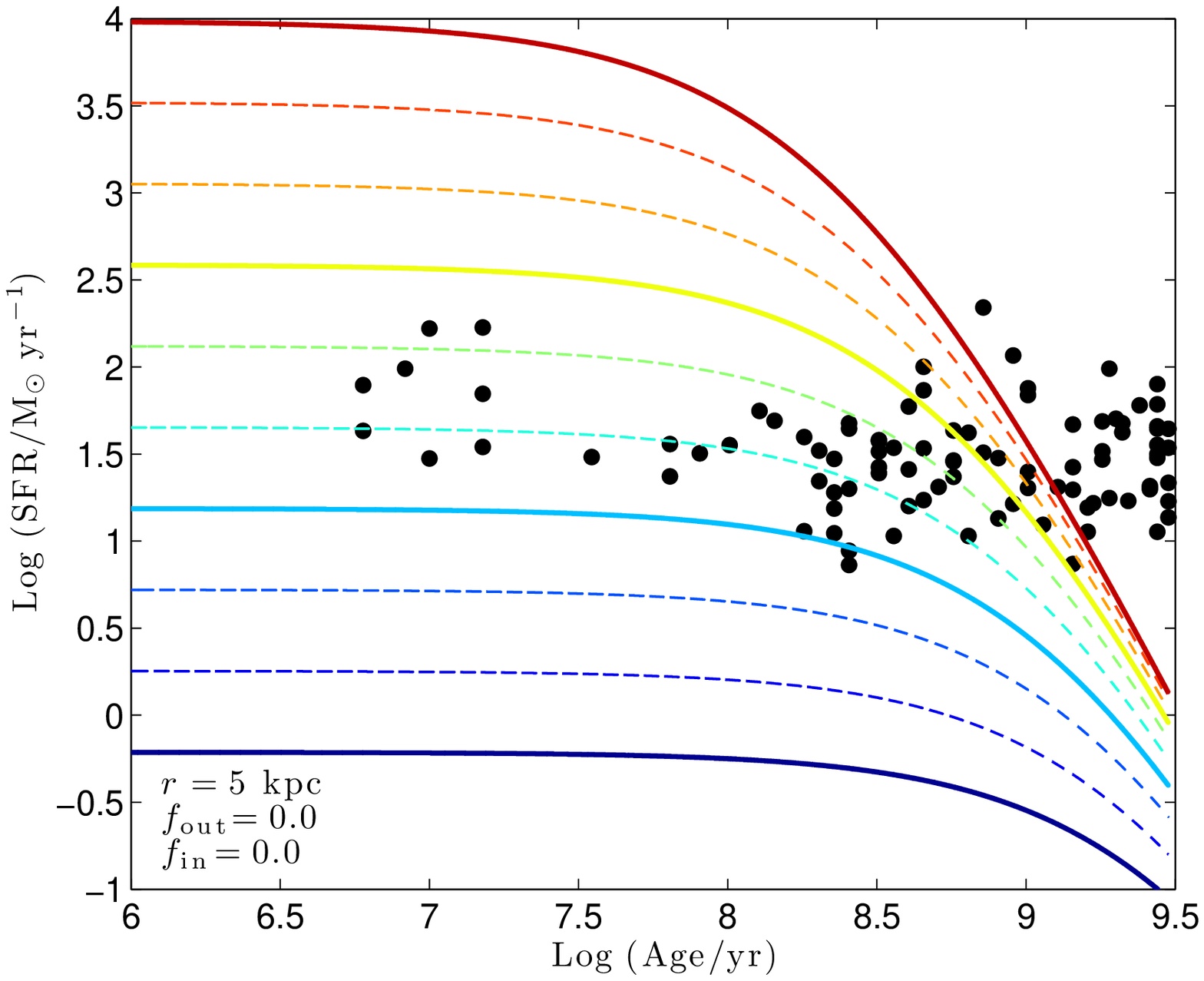}{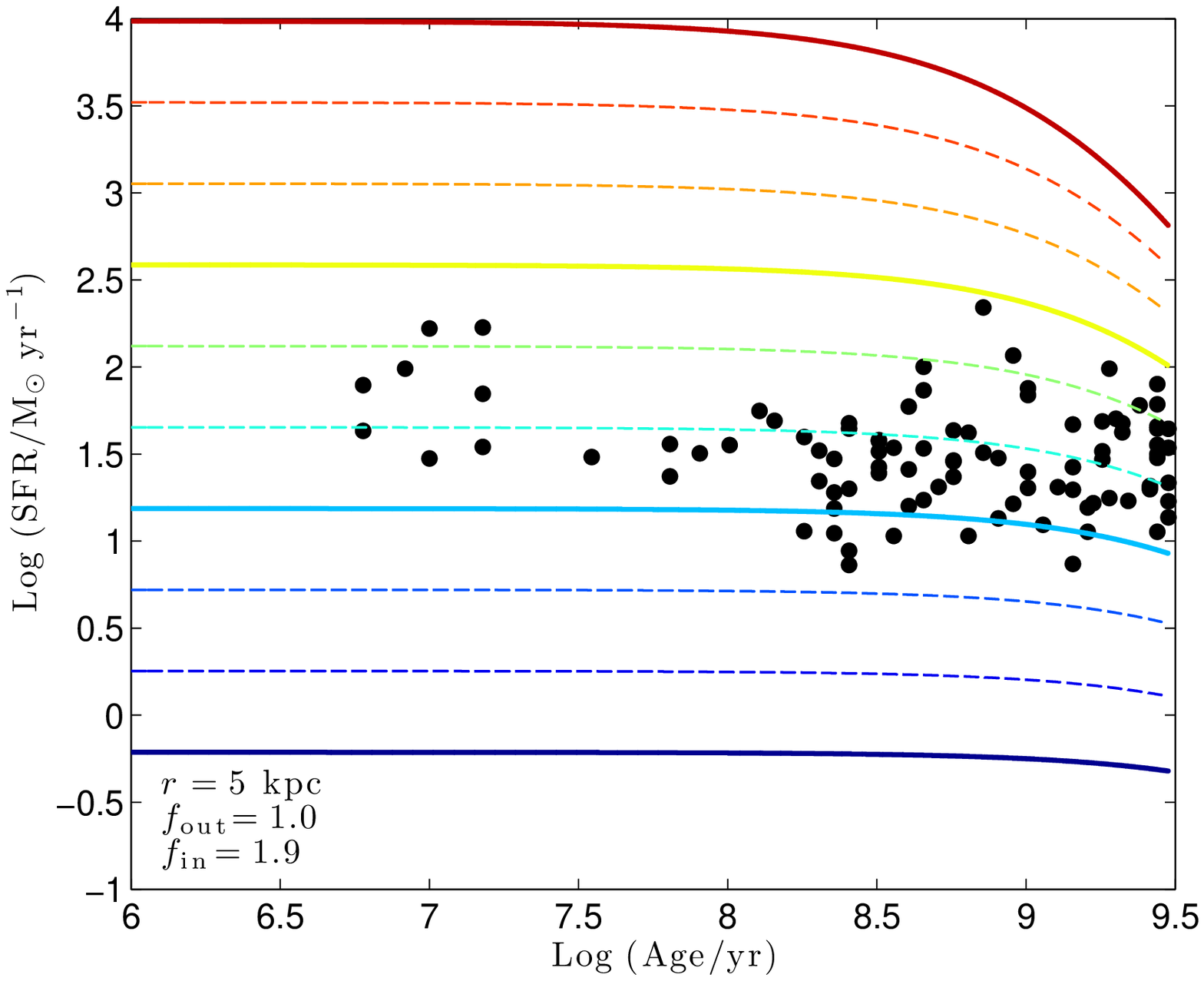}
\caption{The time evolution of the star formation rate as predicted by
  the K-S law for a variety of initial gas masses $M_{\rm i}$.  From
  top to bottom, the solid lines show $M_{\rm i}=10^9$ \msun\ (dark
  blue), $M_{\rm i}=10^{10}$ \msun\ (light blue), $M_{\rm i}=10^{11}$
  \msun\ (yellow), and $M_{\rm i}=10^{12}$ \msun\ (red).  The dashed
  lines show intermediate values.  Data points are from
  \protect\citet{ess+06stars}.  The left panel shows a closed box
  model with no inflows or outflows; the right panel shows an outflow
  rate equal to the star formation rate and a gas accretion rate 1.9
  times the SFR, such that the accretion rate is 95\% of the gas
  processing rate.}
\label{fig:sfrt}
\end{figure*}

The star formation history given by Equation \ref{eq:sfrt} can be
compared to the more familiar exponential star formation history
$\phi(t)\propto e^{-t/\tau}$ by setting the SFRs of the two models
equal at $t_b$ and determining the resulting value of $\tau$.  The
result is $\tau=t_b/2.4$ (270 Myr for the example of
Figure~\ref{fig:ks}), and this exponentially declining SFR is shown by
the red dashed line in Figure~\ref{fig:ks}.  The two star formation
histories are quite similar until $t_b$, when the $\tau$ model begins
to decline more rapidly.

The time evolution of the SFR implied by the K-S law is compared with
observations of \ztwo\ galaxies in Figure~\ref{fig:sfrt}.  The left
panel shows the above model for a range of initial gas masses, from
$M_{\rm i}=10^9$ \msun\ for the bottom blue curve to $M_{\rm
i}=10^{12}$ \msun\ for the red curve at top. The galaxy size is again
fixed at $r=5$ kpc, which is within a factor of $\sim2$ of measured
galaxy sizes at these redshifts; varying $r$ by a factor of two has no
significant impact on the results.  The black points show data from
\citet{ess+06stars}, with star formation rates determined from
extinction-corrected \Ha\ emission and ages from spectral energy
distribution (SED) modeling.  Typical uncertainties are a factor of
$\sim2$ in SFR and a factor of $\sim3$ in age, including uncertainties
in the star formation history; the plotted points assume a constant
star formation history.  These data points are representative of
typical star-forming galaxies at \ztwo; galaxies with significantly lower SFRs
may exist but are too faint to be identified with current
instrumentation, while galaxies with higher SFRs are relatively rare
but can be identified by their strong IR or submillimeter emission
\citep{pmd+06,rsf+06,cbsi05}.  As is usual with the comparison of smooth
model star formation histories and real galaxies, the models can be
assumed to represent either a broad time average of more episodic star
formation events or a smooth, continuous process.  The curves
adequately represent many of the galaxies in the sample, but clearly
do not account for the oldest galaxies, which still show significant
star formation.

Star formation histories of high redshift galaxies are difficult to
constrain, but there is considerable evidence for sustained (either
continuous or episodic) star formation in galaxies at \ztwo, and in
galaxies with ages $\gtrsim1$ Gyr in particular.  Old, massive
(stellar mass $M_{\star}\gtrsim 10^{11}$ \msun) galaxies generally
have the best understood stellar populations; \citet{sse+05} used SED
modeling to show that massive, UV-selected galaxies are best fit with
slowly declining star formation histories with $\tau\sim1$--2 Gyr and
ages of 2--3 Gyr (or by a burst of current star formation superposed
on a maximally old stellar population).  \citet{pmd+06} observed
massive, red ($J-K>2.3$) galaxies, most of which are still forming
stars at a high rate at \ztwo, and found that the earliest star
formation was in the distant past and that most of the stellar mass
had been accumulated well before the time of observation.
\citet{ess+06stars} found from comparisons of \Ha\ SFRs and inferred
ages that the current SFR is a reasonable representation of the past
average for most UV-selected galaxies at \ztwo, but the most massive
galaxies require higher past star formation rates.  Taking a different
approach to star formation histories, \citet{asp+05} used clustering
observations to compare the number densities of UV-selected galaxies
at \ztwo\ and the haloes that can host them, and concluded from the
roughly equal densities that the duty cycle (the fraction of time in
which star formation is ``on'' at a level detectable by the UV
selection criteria) of star formation in these galaxies is of order
unity.

If star formation proceeds according to the K-S law as described
above, no reasonable combination of size and initial gas mass can
sustain star formation over the Gyr timescales required by these
observations.  Moreover, the galaxies have thus far been treated as
closed boxes, which is certainly not the case; strong outflows are
observed to be ubiquitous in star-forming galaxies at these redshifts
(e.g.\ \citealt{pss+01,ssp+03,sci+03}), and observations at low and
high redshifts suggest that the mass outflow rates are comparable to
the star formation rate \citep{m99,psa+00,m03}.  This clearly
exacerbates the gas depletion problem; an outflow rate equal to the
star formation rate effectively halves the lifetime of the galaxy.

The obvious solution is the accretion of additional gas, either in a
continuous flow or through minor mergers.  Such accretion is
theoretically expected at these redshifts (see \S\ref{sec:discuss}),
although evidence for it in observed galaxy spectra is so far minimal.
In order to sustain star formation over such protracted timescales,
gas must be replenished at roughly the rate at which it is processed.
This can be addressed analytically by adding a factor $f$ representing
the net gas flow rate to Equation~\ref{eq:gasmass}:
\begin{equation}
M_{g}(t)=M_{\rm i}-(1-f)\int_0^t \phi(t')dt'. 
\label{eq:gasmassf}
\end{equation}
The gas flow rate $f=f_{\rm i}-f_{\rm o}$, where the outflow rate
$f_{\rm o}$ and the inflow rate $f_{\rm i}$ are given as fractions of
the star formation rate.  For $f>0$, there is a net flow of gas into
the galaxy, but for $f<1$ the total gas mass of the system decreases
with time, because of the additional consumption of gas by star
formation.  As above, we find
\begin{equation}
\phi(t)=n^{n/(n-1)} \left[ A^\prime\,t(1-n)+n \phi_0 ^{(1-n)/n}\right]^{n/(1-n)},
\end{equation}
with $A^\prime=-n(1-f)(Cr^{2(1-n)})^{1/n}$ and the initial star
formation rate $\phi(0)=\phi_0=C M_{\rm i}^nr^{2(1-n)}$.  Note that
the only difference between this solution and that of
Equation~\ref{eq:gasmass} is the factor of $(1-f)$ in the $A^\prime$
term.

This solution is shown in the right panel of Figure~\ref{fig:sfrt},
which shows models with the same range of initial gas masses, but this
time incorporating an accretion rate equal to 95\% of the combined
outflow rate and SFR.  This particular solution has $f_{\rm i}=1.9$
and $f_{\rm 0}=1$, but any combination of inflow and outflow rates
with $f=f_{\rm i}-f_{\rm o}\simeq0.9$ will show the same behavior. The
result is a very shallowly declining SFR, well matched to the
observations.

The models do not fit as well with even a small decrease in the
accretion rate.  An accretion rate of 85\% of the gas processing rate
requires most of the oldest galaxies to have $M_{\rm i}>10^{11}$
\msun, and a further decrease to 80\% fails to account for the SFRs of
about half of the oldest galaxies with the range of initial gas masses
considered.  However, a modest reduction in the required accretion
rate results from the inclusion of the gas returned to the ISM by
evolved stars; as noted above, I have neglected this time-dependent
effect in order to make the modeling more tractable, but an estimate
of the mass of gas returned by star formation can be found from
population synthesis models such as those of \citet{bc03}.  Assuming a
\citet{c03} IMF, at an age of $\sim1$ Gyr a galaxy with a constant
star formation rate of 30 \msunyr\ (the average of the \ztwo\
UV-selected sample) has returned a gas mass of $\sim10^{10}$ \msun\ to
the ISM, for an average (but not constant) rate of $\sim10$ \msunyr,
or $\sim 1/3$ of the SFR.  If the outflow rate is roughly equal to the
star formation rate, the required accretion rate is then $\sim 5/6$ of
the gas processing rate.\footnote{If the \citet{s55} IMF is assumed,
the average SFR of the UV-selected sample is $\sim50$ \msunyr, but the
rate of formation of massive stars that have returned material to the
ISM is the same, since the IMFs differ only at the low end.  Thus
the result is similar for a Salpeter IMF.}  Finally, note that the
essential ingredient here is a reservoir of new gas for star
formation; this could be supplied by gas cooling and falling in from
the halo as well as by gas newly accreted from the surrounding
intergalactic medium (IGM).

Thus far we have constrained only the relative values of the inflow
and outflow rates.  Any model in which the accretion rate is
approximately equal to the gas processing rate will satisfy the
requirements imposed by the star formation rates and ages; for
additional constraints on the magnitude of the inflows and outflows
we must turn to measurements of the gas phase metallicity.

\section{Inflows, Outflows and the\\ Mass-Metallicity Relation}
\label{sec:metals}

\citet{esp+06} observed a correlation between stellar mass and gas
phase metallicity in star-forming galaxies at \ztwo, and used the K-S
law to infer the gas masses and gas fractions of the galaxies.  By
fitting simple chemical evolution models to the relationship between
gas fraction and metallicity, they showed that closed boxes and models
with low outflow rates were inadequate to reproduce the data; with
outflows only, a high outflow rate of $\dot M \sim 4\;\times$ SFR was
required.  Given the above results, however, it's clear that such a
model would deplete the galaxies' gas extremely quickly, and that the
effect of gas inflows on metallicity must also be considered.

I assume for simplicity that gas is accreted at a constant fraction
$f_{\rm i}$ of the star formation rate; this can be viewed either as a
continuous process or as the average of many minor events (for a
thorough treatment of chemical evolution due to discrete events of
accretion and star formation, see \citealt{d07}).  The outflow rate
is also considered to be a constant fraction $f_{\rm o}$ of the
SFR. Then the gas mass is given by
\begin{equation}
M_{\rm g}=M_{\rm i}-\alpha M_{\star}+f_{\rm i} M_{\star} - f_{\rm o}
M_{\star}, 
\label{eq:mgas}
\end{equation}
where $\alpha$ is the fraction of mass remaining locked in stars.  The
metal content $Z$ (defined as the fraction by mass of elements heavier
than helium) evolves according to the standard differential equation
\begin{equation}
\frac{d(Z M_{\rm g})}{dM_{\star}}=y\alpha (1-Z)-\alpha Z - f_{\rm o} Z,
\label{eq:metaldiff}
\end{equation}
where $y$ is the true yield, the ratio of the mass of metals produced
and ejected by star formation to the mass locked in long-lived stars
and remnants.  For a thorough discussion of the derivation of this
equation, see \citet{pagel97}; the reviews by \citet{g97} and
\citet{m02} are also useful.  I have assumed that the metallicity of
the inflowing gas is negligible compared to that of the gas in the
galaxy, and that the outflowing gas has the same metallicity as the
gas that remains in the galaxy.  I also assume for simplicity that
$\alpha=1$, i.e.\ the gas returned to the ISM by star formation is
neglected.  This is not always a good assumption; by an age of a few
Gyr, the returned fraction approaches 40\% of the total mass turned
into stars for a Chabrier IMF.\footnote{The returned fraction at this
age is $\sim25$\% for a Salpeter IMF, with its higher fraction of low
mass stars.}  However, most high redshift galaxies are younger than
this, making the effect less significant, and a proper treatment of the
time-varying $\alpha$ is nontrivial.

For $Z\ll1$, a condition which is always true,
Equation~\ref{eq:metaldiff} has the solution (e.g.\ \citealt{ep84})
\begin{equation}
Z=\frac{y\alpha}{f_{\rm i}} \left[ 1- \left( \frac{M_{\rm g}}{M_{\rm
      i}}\right)^{\frac{f_{\rm i}}{\alpha-f_{\rm i}+f_{\rm
        o}}}\right],
\label{eq:z}
\end{equation}
where the ratio of the current to the initial gas mass $M_{\rm g}/M_{\rm i}$
can be written in terms of the gas fraction 
\begin{equation}
\mu=M_{\rm g}/(M_{\rm g}+\alpha M_{\star})
\end{equation}
 as
\begin{equation}
\frac{M_{\rm g}}{M_{\rm i}}=\frac{\mu}{1+(1-\mu)(f_{\rm
    o}/\alpha-f_{\rm i}/\alpha)}.
\label{eq:gasfrac}
\end{equation}

\begin{figure}[t]
\plotone{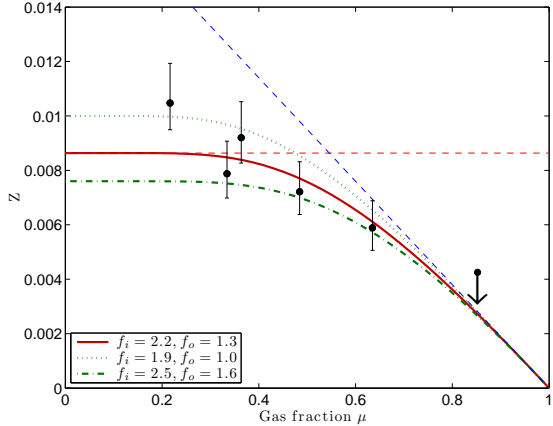}
\caption{The evolution of metallicity with gas fraction as described
  by Equation~\ref{eq:z}.  The data points indicate the metallicities
  and gas fractions found for \ztwo\ galaxies by
  \protect\citet{esp+06}.  Each point represents the average of 14 or
  15 galaxies.  The solid red line is the best fit to the data, and
  the green dotted and dot-dashed lines bracket the range of
  reasonable fits as described in the text.  At high gas fractions $Z$
  rises with decreasing gas fraction as $Z=y \alpha (1-\mu)$ (blue
  dashed line), while at low gas fractions $Z$ approaches the final
  metallicity $Z_{\rm f}=y\alpha/f_{\rm i}$ (red dashed line).  }
\label{fig:metals}
\end{figure}

This result is shown in Figure~\ref{fig:metals}, with the inflow rate
fixed at 95\% of the gas processing rate and a yield
$y=0.019=1.5Z_{\odot}$.\footnote{Using the solar metallicity
$Z_{\odot}=0.0126$ \citep{ags+04}.  Note that the ratio $y/f_{\rm i}$
provides the scaling of the models shown in Figure~\ref{fig:metals},
so a lower yield would also lower the best-fitting accretion rate.}
The solid red line shows the best fit model, with $f_{\rm i}=2.2$ and
$f_{\rm o}=1.3$, while the green dotted and dot-dashed lines bracket
the range of reasonable fits, with $f_{\rm i}=1.9$ and $f_{\rm o}=1$
(the same parameters used in the right panel of Figure~\ref{fig:sfrt}
above) at top and $f_{\rm i}=2.5$ and $f_{\rm o}=1.6$ on the bottom.
It is likely that variations in gas flow rates produce scatter in the
mass-metallicity relation \citep{fd07}, and the green lines may
therefore give some indication of this variation. 

\subsection{Enriched Inflows and Enhanced Outflows}
Thus far I have assumed that the inflowing gas is unenriched, and that
the outflowing gas has the same metallicity as the gas remaining in
the galaxy.  Neither of these conditions are necessarily true; some of
the inflowing gas may be gas expelled by winds falling back onto the
galaxy, and metal-enhanced galactic winds have been observed locally
\citep{mkh02}.  It is reasonably straightforward to incorporate these
additional parameters into the chemical evolution model described
above.  If the metallicity of the inflowing gas is a fraction $z_{\rm
i}$ of the metallicity of the gas in the galaxy, and the metallicity
of the outflowing gas is a fraction $z_{\rm o}$ of the metallicity of
the gas in the galaxy, then Equation~\ref{eq:metaldiff} becomes
\begin{equation}
\frac{d(Z M_{\rm g})}{dM_{\star}}=y\alpha (1-Z)-\alpha Z - f_{\rm
  o}z_{\rm o} Z + f_{\rm i}z_{\rm i} Z.
\label{eq:metaldiff2}
\end{equation}
Combining this with Equation~\ref{eq:mgas} and solving, we find that at
low gas fractions the galaxy reaches a final metallicity
\begin{equation}
Z_{\rm f}=\frac{y \alpha}{f_{\rm i}(1-z_{\rm i})-f_{\rm o}(1-z_{\rm o})},
\label{eq:zf}
\end{equation}
and the evolution of metallicity with gas fraction is described by
\begin{equation}
Z=Z_{\rm f} \left[ 1- \left( \frac{M_{\rm g}}{M_{\rm
      i}}\right)^{\frac{f_{\rm i}(1-z_{\rm i})-f_{\rm o}(1-z_{\rm o})}{\alpha-f_{\rm i}+f_{\rm
        o}}}\right],
\end{equation}
where $M_{\rm g}/M_{\rm i}$ is related to the gas fraction by
Equation~\ref{eq:gasfrac} above.  The final metallicity $Z_{\rm f}$ is
shown by the dashed red line in Figure~\ref{fig:metals}, for the
simple case $z_{\rm i}=0$ and $z_{\rm o}=1$.  It is clear from
Equation~\ref{eq:zf} that increasing the metallicity of the inflowing
gas will also require an increase in the inflow rate in order to
arrive at the observed $Z_{\rm f}\simeq0.01$, while further increasing
the metallicity of the outflows will decrease the inflow and outflow
rates required to reach this same $Z_{\rm f}$ (assuming that the
inflow and outflow rates are related as required by the results of
\S\ref{sec:ks} above).  The two effects may also cancel each other
out; if, for example, $z_{\rm i}=0.5$ and $z_{\rm f}=1.5$, $Z_{\rm
f}\simeq0.01$ is reached with $f_{\rm i}=2.2$ and $f_{\rm o}=1.3$, the
same best-fit parameters shown in Figure~\ref{fig:metals} (although
the shape of this model is slightly different).  In this case the true
accretion rate of new gas is only half that of the $z_{\rm i}=0$ case,
because half of the accreting gas is actually outflow gas returning to
the galaxy.

The addition of two more free parameters clearly decreases our ability
to constrain the gas flow rates; even with the inflow rate fixed at
approximately the gas processing rate, the model now has many
acceptable solutions.  However, given the current lack of constraints
on $z_{\rm i}$ and $z_{\rm o}$, it seems sensible to adopt the simple
model shown in Figure~\ref{fig:metals} for the purposes of further
discussion.  The range of parameter space shown provides an excellent
match to the relationship between metallicity and gas fraction
inferred for the \ztwo\ galaxies, while satisfying the demands of the
K-S law and all available information on gas flow rates.

\section{Discussion}
\label{sec:discuss}

The above results provide a coherent picture in which strong star
formation is sustained by the accretion of gas at approximately the
gas processing rate, the outflow rate is roughly equal to the SFR, and
metal enrichment is modulated by both outflows and inflows.  This is
not a new result; \citet{fd07} reached many of the same conclusions
using cosmological hydrodynamic simulations to reproduce the \ztwo\
mass-metallicity relation, and the idea of a system in which star
formation is balanced by inflow dates to work by \citet{l72} and
\citet{tl78}.  Whatever the methods used to reach these conclusions,
however, many questions remain about the mechanisms of gas flows and
chemical enrichment at high redshift.

The only quantity not yet tied to observations is the gas accretion
rate, which the models require to be approximately equal to the gas
processing rate.  If the outflow rate is roughly equal to the SFR, the
required accretion rate is $\sim60$ \msunyr\ (assuming the average SFR
of the UV-selected sample; \citealt{ess+06stars}), and as much as
several hundred \msunyr\ or higher for the most rapidly star forming
galaxies.  These values are in general agreement with theoretical
expectations.  For example, the predicted average gas accretion rates
for galaxies in $10^{12}$ \msun\ haloes\footnote{Clustering results
indicate that the \ztwo\ UV-selected galaxies are typically associated
with $\sim10^{12}$ \msun\ haloes \citep{asp+05}.}  given by
\citet{kkwd05} are $\sim50$ \msunyr\ at the relevant redshifts, rising
to several hundred \msunyr\ for the $10^{13}$ \msun\ haloes expected
to host the most massive galaxies (though more recent simulations
indicate rates a factor of $\sim2$ or more lower; D. Kere{\v s},
private communication).  Gas accretion and cooling rates from the
semi-analytic models of \citet{csw+06} are also of the right order of
magnitude.

The question remains as to why such high accretion rates have not yet
been observed.  One difficulty is that the velocity range of inflowing
gas is likely to be much narrower than the several hundred \kms\
spread observed in the outflows.  Another suggestion is that cold,
filamentary accretion may dominate at high redshifts
\citep{kkwd05,db06}, in which case detection would depend strongly on
projection effects; alternatively, the accretion could occur largely
in the form of minor mergers.  Another possibility is that the
accreting gas may be too hot to produce signatures in the observed
wavebands, or such signatures may simply be too weak to detect.  Even
if the hot accreting gas does produce \ion{C}{4} emission, for
example, the line would likely be weak because of the low metallicity
of the gas, and it would be superposed on the already complicated
\ion{C}{4} profile.  Detailed modeling of the likely line strengths
would help to place limits on detectable accretion rates.  Gas heated
to the virial temperature of $\sim10^6$ K must also be sufficiently
cooled in order to fuel star formation; work by \citet{yssw02} and
\citet{csw+06}, among others, discusses the mechanisms by which this
might proceed.

The strong star formation and gas accretion discussed herein will not
continue indefinitely.  Observations suggest that most of the
star-forming galaxies currently detected at \ztwo--3 will become
largely passively evolving by $z\sim1$ \citep{asp+05,pmd+06}.  Because
the high observed star formation rates require accretion of new gas to
sustain, a decline in the SFR is a natural consequence of the drop in
accretion rates at lower redshifts predicted by theoretical models.
Many theorists and observers have also proposed that AGN feedback may
be responsible for the termination of star formation
(e.g.\ \citealt{hhc+06} and references therein).  If an additional
mechanism to shut off star formation is required, this is a plausible
candidate, as the ubiquity of outflows suggests that starburst-driven
winds may regulate star formation but do not usually terminate it, and
this work implies that strong accretion and outflows may operate
simultaneously, or at least alternate in relatively quick succession.

Finally, these results underscore the importance of metallicity
measurements for understanding gas flows.  Until the flows can be
observed and quantified directly, measurements of gas phase abundances
offer the best hope for constraining the outflow and inflow rates of
galaxies at high redshift.  There are still considerable difficulties
associated with the measurements of these metallicities, but we look
forward to improved constraints from new IR spectra and
photoionization modeling, and to more direct estimates of outflow
rates from detailed spectra (e.g.\ \citealt{psa+00}).  Such
measurements will give a far more robust picture of star formation,
gas flows and metallicity at high redshift than these simple models
can provide.

\acknowledgements Chuck Steidel, Alice Shapley and Max Pettini were
essential to the acquisition and interpretation of the data on which
this work is based and provided valuable comments.  I would like to
thank the anonymous referee for helpful suggestions, Dusan Kere{\v s}
and Kristian Finlator for useful discussions, and David Kaplan for
technical assistance.  DKE is supported by the CfA Fellowship Program
of the Harvard-Smithsonian Center for Astrophysics.


\begin{thebibliography}{48}
\expandafter\ifx\csname natexlab\endcsname\relax\def\natexlab#1{#1}\fi

\bibitem[{{Adelberger} {et~al.}(2005){Adelberger}, {Steidel}, {Pettini},
  {Shapley}, {Reddy}, \& {Erb}}]{asp+05}
{Adelberger}, K.~L., {Steidel}, C.~C., {Pettini}, M., {Shapley}, A.~E.,
  {Reddy}, N.~A., \& {Erb}, D.~K. 2005, \apj, 619, 697

\bibitem[{{Asplund} {et~al.}(2004){Asplund}, {Grevesse}, {Sauval}, {Allende
  Prieto}, \& {Kiselman}}]{ags+04}
{Asplund}, M., {Grevesse}, N., {Sauval}, A.~J., {Allende Prieto}, C., \&
  {Kiselman}, D. 2004, \aap, 417, 751

\bibitem[{{Baker} {et~al.}(2004){Baker}, {Tacconi}, {Genzel}, {Lehnert}, \&
  {Lutz}}]{btg+04}
{Baker}, A.~J., {Tacconi}, L.~J., {Genzel}, R., {Lehnert}, M.~D., \& {Lutz}, D.
  2004, \apj, 604, 125

\bibitem[{{Bouch{\'e}} {et~al.}(2007){Bouch{\'e}}, {Cresci}, {Davies},
  {Eisenhauer}, {Forster Schreiber}, {Genzel}, {Gillessen}, {Lehnert}, {Lutz},
  {Nesvadba}, {Shapiro}, {Sternberg}, {Tacconi}, {Verma}, {Cimatti}, {Daddi},
  {Renzini}, {Erb}, {Shapley}, \& {Steidel}}]{bcd+07}
{Bouch{\'e}}, N., {Cresci}, G., {Davies}, R., {Eisenhauer}, F., {Forster
  Schreiber}, N.~M., {Genzel}, R., {Gillessen}, S., {Lehnert}, M., {Lutz}, D.,
  {Nesvadba}, N., {Shapiro}, K.~L., {Sternberg}, A., {Tacconi}, L.~J., {Verma},
  A., {Cimatti}, A., {Daddi}, E., {Renzini}, A., {Erb}, D.~K., {Shapley}, A.,
  \& {Steidel}, C.~C. 2007, ArXiv e-prints, 0706.2656, \apj, in press

\bibitem[{{Bruzual} \& {Charlot}(2003)}]{bc03}
{Bruzual}, G. \& {Charlot}, S. 2003, \mnras, 344, 1000

\bibitem[{{Calura} {et~al.}(2007){Calura}, {Jimenez}, {Panter}, {Matteucci}, \&
  {Heavens}}]{cjp+07}
{Calura}, F., {Jimenez}, R., {Panter}, B., {Matteucci}, F., \& {Heavens}, A.~F.
  2007, ArXiv e-prints, 0707.1345

\bibitem[{{Chabrier}(2003)}]{c03}
{Chabrier}, G. 2003, \pasp, 115, 763

\bibitem[{{Chapman} {et~al.}(2005){Chapman}, {Blain}, {Smail}, \&
  {Ivison}}]{cbsi05}
{Chapman}, S.~C., {Blain}, A.~W., {Smail}, I., \& {Ivison}, R.~J. 2005, \apj,
  622, 772

\bibitem[{{Coppin} {et~al.}(2007){Coppin}, {Swinbank}, {Neri}, {Cox}, {Smail},
  {Ellis}, {Geach}, {Siana}, {Teplitz}, {Dye}, {Kneib}, {Edge}, \&
  {Richard}}]{csn+07}
{Coppin}, K.~E.~K., {Swinbank}, A.~M., {Neri}, R., {Cox}, P., {Smail}, I.,
  {Ellis}, R.~S., {Geach}, J.~E., {Siana}, B., {Teplitz}, H., {Dye}, S.,
  {Kneib}, J.~., {Edge}, A.~C., \& {Richard}, J. 2007, ArXiv e-prints,
  0705.1721

\bibitem[{{Croton} {et~al.}(2006){Croton}, {Springel}, {White}, {De Lucia},
  {Frenk}, {Gao}, {Jenkins}, {Kauffmann}, {Navarro}, \& {Yoshida}}]{csw+06}
{Croton}, D.~J., {Springel}, V., {White}, S.~D.~M., {De Lucia}, G., {Frenk},
  C.~S., {Gao}, L., {Jenkins}, A., {Kauffmann}, G., {Navarro}, J.~F., \&
  {Yoshida}, N. 2006, \mnras, 367, 864

\bibitem[{{Dalcanton}(2007)}]{d07}
{Dalcanton}, J.~J. 2007, \apj, 658, 941

\bibitem[{{Dekel} \& {Birnboim}(2006)}]{db06}
{Dekel}, A. \& {Birnboim}, Y. 2006, \mnras, 368, 2

\bibitem[{{Edmunds} \& {Pagel}(1984)}]{ep84}
{Edmunds}, M.~G. \& {Pagel}, B.~E.~J. 1984, in {\it Stellar Nucleosynthesis},
  ed. C.~{Chiosi} \& A.~{Renzini}, 341

\bibitem[{{Erb} {et~al.}(2006{\natexlab{a}}){Erb}, {Shapley}, {Pettini},
  {Steidel}, {Reddy}, \& {Adelberger}}]{esp+06}
{Erb}, D.~K., {Shapley}, A.~E., {Pettini}, M., {Steidel}, C.~C., {Reddy},
  N.~A., \& {Adelberger}, K.~L. 2006{\natexlab{a}}, \apj, 644, 813

\bibitem[{{Erb} {et~al.}(2006{\natexlab{b}}){Erb}, {Steidel}, {Shapley},
  {Pettini}, {Reddy}, \& {Adelberger}}]{ess+06stars}
{Erb}, D.~K., {Steidel}, C.~C., {Shapley}, A.~E., {Pettini}, M., {Reddy},
  N.~A., \& {Adelberger}, K.~L. 2006{\natexlab{b}}, \apj, 647, 128

\bibitem[{{Erb} {et~al.}(2006{\natexlab{c}}){Erb}, {Steidel}, {Shapley},
  {Pettini}, {Reddy}, \& {Adelberger}}]{ess+06mass}
---. 2006{\natexlab{c}}, \apj, 646, 107

\bibitem[{{Finlator} \& {Dav{\'e}}(2007)}]{fd07}
{Finlator}, K. \& {Dav{\'e}}, R. 2007, ArXiv e-prints, 0704.3100

\bibitem[{{Gibson}(1997)}]{g97}
{Gibson}, B.~K. 1997, \mnras, 290, 471

\bibitem[{{Hopkins} {et~al.}(2006){Hopkins}, {Hernquist}, {Cox}, {Di Matteo},
  {Robertson}, \& {Springel}}]{hhc+06}
{Hopkins}, P.~F., {Hernquist}, L., {Cox}, T.~J., {Di Matteo}, T., {Robertson},
  B., \& {Springel}, V. 2006, \apjs, 163, 1

\bibitem[{{Juneau} {et~al.}(2005){Juneau}, {Glazebrook}, {Crampton},
  {McCarthy}, {Savaglio}, {Abraham}, {Carlberg}, {Chen}, {Le Borgne}, {Marzke},
  {Roth}, {J{\o}rgensen}, {Hook}, \& {Murowinski}}]{jgc+05}
{Juneau}, S., {Glazebrook}, K., {Crampton}, D., {McCarthy}, P.~J., {Savaglio},
  S., {Abraham}, R., {Carlberg}, R.~G., {Chen}, H., {Le Borgne}, D., {Marzke},
  R.~O., {Roth}, K., {J{\o}rgensen}, I., {Hook}, I., \& {Murowinski}, R. 2005,
  \apjl, 619, L135

\bibitem[{{Kennicutt}(1998)}]{k98schmidt}
{Kennicutt}, R.~C. 1998, \apj, 498, 541

\bibitem[{{Kere{\v s}} {et~al.}(2005){Kere{\v s}}, {Katz}, {Weinberg}, \&
  {Dav{\'e}}}]{kkwd05}
{Kere{\v s}}, D., {Katz}, N., {Weinberg}, D.~H., \& {Dav{\'e}}, R. 2005,
  \mnras, 363, 2

\bibitem[{{Kobulnicky} \& {Koo}(2000)}]{kk00}
{Kobulnicky}, H.~A. \& {Koo}, D.~C. 2000, \apj, 545, 712

\bibitem[{{K{\"o}ppen} \& {Edmunds}(1999)}]{ke99}
{K{\"o}ppen}, J. \& {Edmunds}, M.~G. 1999, \mnras, 306, 317

\bibitem[{{K{\"o}ppen} \& {Hensler}(2005)}]{kh05}
{K{\"o}ppen}, J. \& {Hensler}, G. 2005, \aap, 434, 531

\bibitem[{{Larson}(1972)}]{l72}
{Larson}, R.~B. 1972, \nat, 236, 21

\bibitem[{{Maier} {et~al.}(2006){Maier}, {Lilly}, {Carollo}, {Meisenheimer},
  {Hippelein}, \& {Stockton}}]{mlc+06}
{Maier}, C., {Lilly}, S.~J., {Carollo}, C.~M., {Meisenheimer}, K., {Hippelein},
  H., \& {Stockton}, A. 2006, \apj, 639, 858

\bibitem[{{Martin}(1999)}]{m99}
{Martin}, C.~L. 1999, \apj, 513, 156

\bibitem[{{Martin}(2003)}]{m03}
{Martin}, C.~L. 2003, in Revista Mexicana de Astronomia y Astrofisica
  Conference Series, 56--59

\bibitem[{{Martin} {et~al.}(2002){Martin}, {Kobulnicky}, \& {Heckman}}]{mkh02}
{Martin}, C.~L., {Kobulnicky}, H.~A., \& {Heckman}, T.~M. 2002, \apj, 574, 663

\bibitem[{{Matteucci}(2002)}]{m02}
{Matteucci}, F. 2002, ArXiv Astrophysics e-prints

\bibitem[{{Murray} {et~al.}(2005){Murray}, {Quataert}, \& {Thompson}}]{mqt05}
{Murray}, N., {Quataert}, E., \& {Thompson}, T.~A. 2005, \apj, 618, 569

\bibitem[{{Pagel}(1997)}]{pagel97}
{Pagel}, B.~E.~J. 1997, {Nucleosynthesis and Chemical Evolution of Galaxies}
  (Nucleosynthesis and Chemical Evolution of Galaxies, by Bernard E.~J.~Pagel,
  pp.~392.~ISBN 0521550610.~Cambridge, UK: Cambridge University Press, October
  1997.)

\bibitem[{{Panter} {et~al.}(2007){Panter}, {Jimenez}, {Heavens}, \&
  {Charlot}}]{pjhc07}
{Panter}, B., {Jimenez}, R., {Heavens}, A.~F., \& {Charlot}, S. 2007, \mnras,
  378, 1550

\bibitem[{{Papovich} {et~al.}(2006){Papovich}, {Moustakas}, {Dickinson}, {Le
  Floc'h}, {Rieke}, {Daddi}, {Alexander}, {Bauer}, {Brandt}, {Dahlen}, {Egami},
  {Eisenhardt}, {Elbaz}, {Ferguson}, {Giavalisco}, {Lucas}, {Mobasher},
  {P{\'e}rez-Gonz{\'a}lez}, {Stutz}, {Rieke}, \& {Yan}}]{pmd+06}
{Papovich}, C., {Moustakas}, L.~A., {Dickinson}, M., {Le Floc'h}, E., {Rieke},
  G.~H., {Daddi}, E., {Alexander}, D.~M., {Bauer}, F., {Brandt}, W.~N.,
  {Dahlen}, T., {Egami}, E., {Eisenhardt}, P., {Elbaz}, D., {Ferguson}, H.~C.,
  {Giavalisco}, M., {Lucas}, R.~A., {Mobasher}, B., {P{\'e}rez-Gonz{\'a}lez},
  P.~G., {Stutz}, A., {Rieke}, M.~J., \& {Yan}, H. 2006, \apj, 640, 92

\bibitem[{{Pettini} {et~al.}(2001){Pettini}, {Shapley}, {Steidel}, {Cuby},
  {Dickinson}, {Moorwood}, {Adelberger}, \& {Giavalisco}}]{pss+01}
{Pettini}, M., {Shapley}, A.~E., {Steidel}, C.~C., {Cuby}, J., {Dickinson}, M.,
  {Moorwood}, A.~F.~M., {Adelberger}, K.~L., \& {Giavalisco}, M. 2001, \apj,
  554, 981

\bibitem[{{Pettini} {et~al.}(2000){Pettini}, {Steidel}, {Adelberger},
  {Dickinson}, \& {Giavalisco}}]{psa+00}
{Pettini}, M., {Steidel}, C.~C., {Adelberger}, K.~L., {Dickinson}, M., \&
  {Giavalisco}, M. 2000, \apj, 528, 96

\bibitem[{{Reddy} {et~al.}(2006){Reddy}, {Steidel}, {Fadda}, {Yan}, {Pettini},
  {Shapley}, {Erb}, \& {Adelberger}}]{rsf+06}
{Reddy}, N.~A., {Steidel}, C.~C., {Fadda}, D., {Yan}, L., {Pettini}, M.,
  {Shapley}, A.~E., {Erb}, D.~K., \& {Adelberger}, K.~L. 2006, \apj, 644, 792

\bibitem[{{Salpeter}(1955)}]{s55}
{Salpeter}, E.~E. 1955, \apj, 121, 161

\bibitem[{{Savaglio} {et~al.}(2004){Savaglio}, {Glazebrook}, {Abraham},
  {Crampton}, {Chen}, {McCarthy}, {J{\o}rgensen}, {Roth}, {Hook}, {Marzke},
  {Murowinski}, \& {Carlberg}}]{sga+04}
{Savaglio}, S., {Glazebrook}, K., {Abraham}, R.~G., {Crampton}, D., {Chen},
  H.-W., {McCarthy}, P.~J.~P., {J{\o}rgensen}, I., {Roth}, K.~C., {Hook},
  I.~M., {Marzke}, R.~O., {Murowinski}, R.~G., \& {Carlberg}, R.~G. 2004, \apj,
  602, 51

\bibitem[{{Schmidt}(1963)}]{s63}
{Schmidt}, M. 1963, \apj, 137, 758

\bibitem[{{Shapley} {et~al.}(2004){Shapley}, {Erb}, {Pettini}, {Steidel}, \&
  {Adelberger}}]{sep+04}
{Shapley}, A.~E., {Erb}, D.~K., {Pettini}, M., {Steidel}, C.~C., \&
  {Adelberger}, K.~L. 2004, \apj, 612, 108

\bibitem[{{Shapley} {et~al.}(2005){Shapley}, {Steidel}, {Erb}, {Reddy},
  {Adelberger}, {Pettini}, {Barmby}, \& {Huang}}]{sse+05}
{Shapley}, A.~E., {Steidel}, C.~C., {Erb}, D.~K., {Reddy}, N.~A., {Adelberger},
  K.~L., {Pettini}, M., {Barmby}, P., \& {Huang}, J. 2005, \apj, 626, 698

\bibitem[{{Shapley} {et~al.}(2003){Shapley}, {Steidel}, {Pettini}, \&
  {Adelberger}}]{ssp+03}
{Shapley}, A.~E., {Steidel}, C.~C., {Pettini}, M., \& {Adelberger}, K.~L. 2003,
  \apj, 588, 65

\bibitem[{{Smail} {et~al.}(2003){Smail}, {Chapman}, {Ivison}, {Blain},
  {Takata}, {Heckman}, {Dunlop}, \& {Sekiguchi}}]{sci+03}
{Smail}, I., {Chapman}, S.~C., {Ivison}, R.~J., {Blain}, A.~W., {Takata}, T.,
  {Heckman}, T.~M., {Dunlop}, J.~S., \& {Sekiguchi}, K. 2003, \mnras, 342, 1185

\bibitem[{{Tinsley}(1980)}]{t80}
{Tinsley}, B.~M. 1980, Fundamentals of Cosmic Physics, 5, 287

\bibitem[{{Tinsley} \& {Larson}(1978)}]{tl78}
{Tinsley}, B.~M. \& {Larson}, R.~B. 1978, \apj, 221, 554

\bibitem[{{Yoshida} {et~al.}(2002){Yoshida}, {Stoehr}, {Springel}, \&
  {White}}]{yssw02}
{Yoshida}, N., {Stoehr}, F., {Springel}, V., \& {White}, S.~D.~M. 2002, \mnras,
  335, 762

\end{thebibliography}
\end{document}